\documentclass[aps,prb,showpacs,twoside,twocolumn,10pt]{revtex4-1}
\usepackage[colorlinks=true, citecolor=red, urlcolor=blue ]{hyperref}
\usepackage{times,epsfig,amssymb,amsfonts,amsmath,bm,subfigure,mathtools,amsthm,braket,soul,enumitem,xcolor,physics,graphics,graphicx}
\UseRawInputEncoding
\usepackage[normalem]{ulem}
\usepackage{comment}
\usepackage{epstopdf}

\usepackage{float}

\begin{document}

\title{MagGen: A graph aided deep generative model for inverse design of stable, permanent magnets
}

\author{Sourav Mal$^{1}$, Gaurav Seal$^{2}$, Prasenjit Sen$^{1,3}$}
\affiliation{$^1$Harish-Chandra Research Institute, a CI of Homi
Bhabha National Institute, Chhatnag Road, Jhunsi, Prayagraj 211019,
India}
\affiliation{$^2$ Indian Institute of Science Education and Research Thiruvananthapuram, Thiruvananthapuram, Kerala 695551, India}
\affiliation{$^3$ Department of Physics, Indian Institute of Science Education and Research, Tirupati, Karkambadi Road, Tirupati 517507, India}

\begin{abstract}
A significant development towards inverse design of materials with well-defined target properties is reported. 
A deep generative model based on variational autoencoder (VAE), conditioned simultaneously by two target properties, is developed to inverse design stable magnetic materials.
Structure of the physics informed, property embedded latent space of the model is analyzed using graph theory, based on the idea of similarity index. The graph idea is shown to be useful for generating new materials that are likely to satisfy target properties. 
An impressive $\sim 96\%$ of the generated materials is found to satisfy the target properties as per predictions from the target learning branches.
This is a huge improvement over approaches that do not condition the VAE latent space by target properties, or do not consider connectivity
of the parent materials perturbing which the new materials are generated. In such models, the fraction of materials satisfying
targets can be as low as $\sim 5\%$. This impressive feat is achieved using a simple real-space only representation called 
Invertible Real-space Crystallographic Representation (IRCR), that can
be directly read from material cif files. Model predictions are finally validated by performing DFT
calculations on a randomly chosen subset of materials. Performance of the present model using IRCR is comparable or superior
to that of the models reported earlier. This model for magnetic material generation, MagGen, is applied to the problem of 
designing rare earth free permanent magnets with promising results.

\end{abstract}

\maketitle

A key challenge in materials research is to design new materials with desired properties. In the recent past, high-throughput virtual screening 
(HTVS)~\cite{htvs1,htvs2,htvs3,htvs4,htvs5,htvs6,htvs7,htvs8,htvs9,htvs10} 
has achieved some success in discovering new materials. HTVS screens an initial pool of materials, 
usually generated via chemical intuition or other heuristic methods,
using density functional theory (DFT) calculations in a high throughput mode. More recently, machine learning (ML) based property prediction models~\cite{CGCNN,MEGNet} have greatly accelerated this screening step. 
Despite its successes, the method is limited to explore only a small subspace of the material space as represented
by the initial pool.

Deep learning based generative ML models (GM) have made possible efficient exploration of the vast material space 
enabling the long-standing goal of {\em inverse design}. Inverse design in essence means inverting the usual 
structure and composition to property mapping, and deriving structure and composition given a desired set of properties.

GMs create a continuous representation of materials in a space called the latent space. This can be embedded with properties during the 
training process, and new materials can be generated by sampling points from this space. Although GMs have been successful in inverse 
designing molecules~\cite{doi:10.1021/acscentsci.7b00572,doi:10.1021/acs.molpharmaceut.7b00346,doi:10.1021/acscentsci.7b00512,doi:10.1021/acs.jcim.7b00690}, 
they face considerable challenge in case of crystalline materials due to the absence of an invertible representation of the latter. 
The two most commonly used GMs in inverse design of crystalline materials are variational autoencoder (VAE)~\cite{VAE1,VAE2}, 
and generative adversarial network (GAN)~\cite{GAN}.

3D voxel representation of crystalline materials was used to generate stable $\text{V}_x\text{O}_y$ materials~\cite{iMatGen}, Bi-Se 
materials~\cite{DCGAN} and zeolites~\cite{ZeoGAN}. 
Voxel representation based on 3D electron density maps was used to generate cubic materials~\cite{ACS-JCIM} and magnetic materials for magnetocaloric applications~\cite{Court-ChemMat}. 
Choubisa et al.~\cite{CSFE} used crystal 
site feature embedding (CSFE) representation to inverse design perovskite materials using a VAE. An alternative to voxel image is 2D point cloud representation.
This has been used to inverse design Mg-Mn-O~\cite{CCCG} materials for photoanode applications, and stable ternary materials~\cite{CrystalGAN} using GAN. 
Cubic materials have been designed using CubicGAN~\cite{CubicGAN}, built using a GAN based on a 2D tensor representation. 
In spite of their success in specific situations,
all these applications are restricted to encode either materials with a fixed composition~\cite{iMatGen,DCGAN,ZeoGAN,CrystalGAN, CCCG} 
or those with a fixed crystal structure~\cite{CubicGAN,CSFE,ACS-JCIM,Court-ChemMat}. Moreover, training generative models with voxel representation is highly memory intensive.
Also, it will be extremely challenging to extend Noh et al's~\cite{iMatGen} approach beyond a fixed binary phase.        

Ren et al.~\cite{FTCP} proposed an invertible point cloud representation called Fourier transformed crystal properties (FTCP), 
which can encode crystalline materials with different compositions and structures using a combination of real space and reciprocal space information. 
Computationally, FTCP was a major improvement over voxel representation, as it allows generating materials with varying
structure and composition, and is computationally much less demanding.
It was used to inverse design inorganic materials with user defined formation energies, 
bandgap, thermoelectric power factor and combinations thereof. 
Zhao et al.~\cite{PGCGM} proposed a deep learning based Physics Guided Crystal Generative model (PGCGM) for
designing materials with high structural diversity and symmetry. PGCGM uses 2D tensor representation that combines space group 
affine transformation and a self-augmentation method.

In a significant development beyond FTCP, we present an efficient generative model using only real space features. 
We term this representation as {\em invertible real space crystallographic representation} (IRCR). In this work, IRCR is
used to inverse design stable ferromagnetic materials with large saturation magnetization ($M_s$). 
Design of materials with large saturation magnetization and magnetic anisotropy energy (AE), particularly those without
rare earth elements, is a challenging and important problem~\cite{coey}.
The target material properties are $M_s \ge 1$T~\cite{novamag-paper}, AE should be uniaxial with the anisotropy constant
$K_1 > 1$~MJ/m$^3$~\cite{novamag-paper}. Idea of the anisotropy constant is discussed in section S1 in the Supplementary Information (SI).
Design of such materials has been addressed by us~\cite{JMMM} and other authors~\cite{JMCC,Liao,halder,novomag-paper,novamag-paper,vishina,VISHINA2023119348,zhao-PM,PhysRevMaterials.7.044405}. 
These works used either high-throughput DFT calculations, or ML-assisted screening to identify candidate
materials, followed by validation with DFT calculations. 
So far, generative models have not been employed to address the problem, primarily due to paucity of materials data for AE.
Given this constraint, we aim to design stable ferromagnetic materials using a GM,
and subsequently screen these for AE via DFT. We deliberately set a lower threshold for $M_s$, $M_s \ge 0.5$~T, during the generation process
for convenience in this first magnetic materials design model with magnetization as one of the targets. But as we present later, we do generate 
materials having $M_s > 1$~T that also satisfy the AE target.

We use conditional VAE (cVAE) as the generative model in which the latent space is conditioned by 
formation energy ($h_\text{form}$) and $M_s$ simultaneously. 
$h_\text{form}$ is used as a measure of stability with respect to the isolated constituents. 
Distance from the convex hull ($E_{\rm hull}$) as a measure of stability is calculated only for the newly designed materials. We call this framework for stable, magnetic 
materials generation as {\bf MagGen}. To the best of our knowledge, this is the first attempt to inverse design permanent
magnet materials using any GM, and only the second work after Ref.~\cite{FTCP} attempting to organize the latent space with more than one target properties.      

\begin{figure}
    \centering
    \includegraphics[scale=0.22]{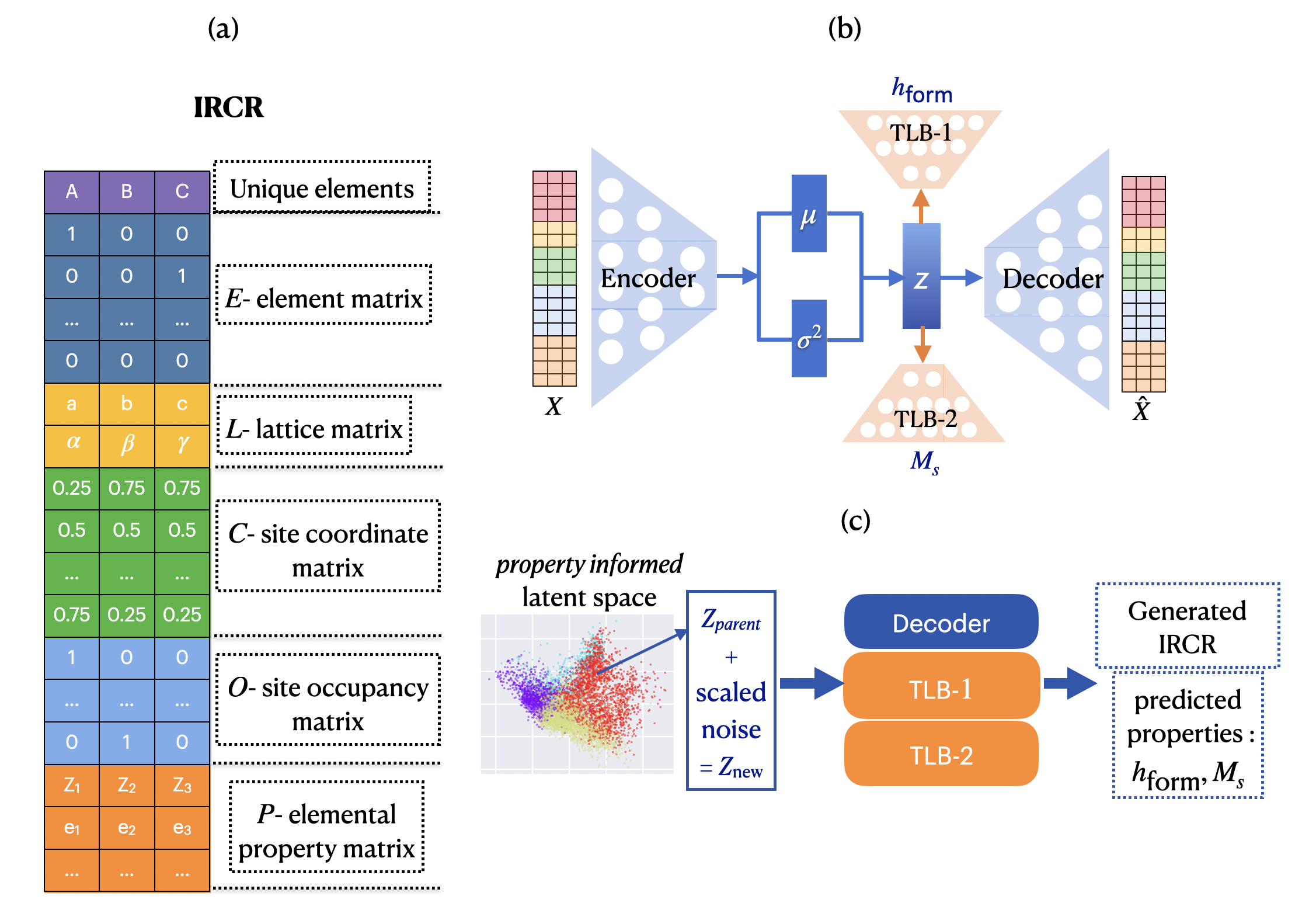}
    \caption{(a) Invertible real space crystallographic representation (IRCR) based on only real space features, (b) The schematic diagram of MagGen consisting of an encoder, a decoder and two target-learning branches, (c) Scheme for new material generation with predicted $h_\text{form}$ and $M_s$. }
    \label{schematic}
\end{figure}

IRCR encodes the structure, composition and elemental properties of bulk crystalline materials. It consists of five matrices, concatenated vertically.
\begin{equation}
    \text{IRCR} = \{E,L,C,O,P\},
\end{equation}
in which (i) {\em E} is the element matrix where each column is a one-hot encoded vector to represent each unique element present in the material. 
We consider up to ternary materials in this work. This fixes the shape of this matrix to ($Z_\text{max}\times3$), where $Z_\text{max}$ is the highest atomic number 
among the elements present in the entire database.    \\
(ii) {\em L} is a ($2\times3$) matrix which encodes the lattice constants a, b, c, and the lattice angles $\alpha$, $\beta$ and $\gamma$ in the two rows. \\
(iii) {\em C} is the site coordinate matrix that contains the fractional coordinates of each lattice site in the unit cell. The shape of this matrix is ($n_\text{sites}\times3$), where 
$n_\text{sites}$ is the maximum number of atoms in the unit cell among all the materials. For materials with number of sites less than 
$n_\text{sites}$, zero padding is used for the empty sites.   \\
(iv) {\em O} is the ($n_\text{sites}\times3$) site occupancy matrix to encode the elements at each lattice site. Each row is one-hot encoded to represent the elements. 
There is a one-to-one correspondence between {\em E} and {\em O} matrices.   \\
(v) {\em P} is the $(6\times 3)$ elemental property matrix. Each column of $P$ represents the property vector of the respective element. 
We use six elemental properties to encode chemical information. These are atomic number ($Z$),  electronegativity ($e$), 
period number ($p$), group number ($g$), its atomic fraction ($s$) in the material and number of d electrons ($n_d$).  The last one is included
as it helps in prediction of magnetic properties~\cite{JMCC,JMMM}. \\
A schematic diagram of IRCR is given in Fig.~\ref{schematic}(a). 

A schematic diagram of MagGen is shown in Fig.~\ref{schematic}(b). It consists of an encoder, 
a decoder and two target-learning branches (TLBs). The encoder uses IRCR as input ($X$) and maps it into an isotropic Gaussian distribution 
$\mathcal{N}(\mu,\sigma^2)$ in a lower dimensional latent space. A point $z$ is sampled from the latent space according to Eq.(\ref{eq:latentpoint}).
\begin{equation}
    z = \mu + \sigma\odot\epsilon, \epsilon \sim \mathcal{N}(0,I),
    \label{eq:latentpoint}
\end{equation}
where $\epsilon$ is a random noise sampled from a unit Gaussian, and $\odot$ denotes element-wise multiplication. 
The decoder maps $z$ back to the original input, called reconstructed output ($\hat{X}$).   

The latent space is regularized by enforcing the encoding distributions to follow a unit Gaussian prior. Any deviation from the prior is expressed as the
Kullback-Leibler (KL) divergence between the encoding distribution and the prior. It enforces the encoded distributions to overlap as much as possible, 
centered around the origin, and ensures creation of a continuous latent space~\cite{VAE1,VAE2}.

The two TLBs map the latent points to the physical properties of interest. 
The branches predicting $h_\text{form}$ and $M_s$ are termed TLB-1 and TLB-2 respectively. 
This combination of regularization and property embedding builds in a structure in the latent space which 
ensures that latent points close to each other lead to similar properties, as established in
greater detail later. We call this space the {\em property informed continuous latent space}.

Since IRCR is a 2D matrix, it can be thought of as an image representing crystalline materials, with each matrix element being the corresponding pixel value. 
From this perspective, we leverage convolutional neural network (CNN)~\cite{CNN}, specifically designed for image processing tasks, to extract meaningful 
features from IRCR. The encoder and the decoder are built using 1D CNNs, and 
the TLBs are built using fully-connected feedforward neural networks (NN). The detailed architecture is discussed in section S2 in SI. 
The network architecture is built using the Tensorflow~\cite{Tensorflow} python library. The encoder, the decoder and the two TLBs 
are trained jointly by minimizing a loss function given in Eq.(\ref{eq:loss}), using the back-propagation~\cite{backprop} algorithm. 
\begin{equation}
    \mathcal{L} = \mathcal{L}_\text{recons} + \alpha_1\mathcal{L}_\text{KL} + \alpha_2\mathcal{L}_{h_\text{form}} + \alpha_3\mathcal{L}_{M_s} .
    \label{eq:loss}
\end{equation}

\noindent Here $\mathcal{L}_\text{recons}$ is the reconstruction loss, expressed by mean squared error between the actual input ($X$) and 
the reconstructed output ($\hat{X}$).  $\mathcal{L}_\text{KL}$ is the regularization term, expressed by the KL divergence between 
the encoding and prior distributions. It has a closed form expression given by Eq.(\ref{eq:KL-loss}).
\begin{equation}
    \mathcal{L_\text{KL}} = D_{\text{KL}} (\mathcal{N}(\mu,\sigma^2)   ||   \mathcal{N}(0,I) )  
    = -\frac{1}{2}\sum[1+log(\sigma^2)-\mu^2-\sigma^2] .
    \label{eq:KL-loss}
\end{equation}
In this work, dimension of the latent space is chosen to be 256. So the sum in Eqn.~\ref{eq:KL-loss} is taken over these 256 dimensions. 
$\mathcal{L}_{h_\text{form}}$ and $\mathcal{L}_{M_s}$ are the mean squared errors between the actual and predicted $h_\text{form}$ and $M_s$ values. 
$\alpha_1$, $\alpha_2$ and $\alpha_3$ are three hyperparameters of the model which need to be optimized prior to training. 
Together they control the trade-off between different terms in the loss function. 

The data set to train MagGen includes inorganic bulk materials taken from three publicly available databases - the Materials Project(MP)~\cite{MP}, 
novomag~\cite{novomag-paper} and Novamag~\cite{novamag-paper}. DFT computed materials properties are listed in these databases. 
In our work, we include up to ternary materials with the number of sites in the unit cell ($n_\text{sites}$) up to 20, and lattice constants less than or equal to 25~\AA. 
But the MagGen framework is not limited to these choices, and can be extended beyond these limits if required. 

After removing the duplicate entries, removing entries that show `NAN' for $h_\text{form}$, the final data set contains 40048 materials. 
Out of these, 37592 are from MP, 1357 from novomag, and 1099 are from Novamag. The data set contains 
materials covering 7 lattice systems, 177 space groups and 86 different elements up to Pu ($Z=94$). $1.2\%$ of the materials are elemental, 
$31\%$ are binary and $67.8\%$ are ternary. $74\%$ of the materials are stable having $h_\text{form}\le0$, and only $34.4\%$ are magnetic which have $M_s\ge 0.1$~T.  

The data set is split in an 80:20 ratio to create training and test sets. For tuning the hyperparameters, 10\% of the training set is used as validation set, 
and after optimizing the hyperparameters, the whole training set is used to train MagGen with the optimal set of hyperparameters. 
Finally, the model performance is assessed on the test set. The architecture of MagGen and the optimal hyperparameters are discussed in section S2 in SI. 

The performance of MagGen is assessed by analyzing the reconstruction performance of the decoder, property prediction accuracy of the 
two TLBs, and the organization of the latent space for the test set materials. The first two are summarized in Table~\ref{cVAE}. 
We compare our results with those of Ren et al.~\cite{FTCP} wherever they overlap.

\begin{table}[ht]
    \centering
     \caption{ Reconstruction performance of the decoder and property prediction performances of the two target-learning branches of MagGen using IRCR. The similar quantities for FTCP are mentioned for comparison.
     }
     \vspace{0.2cm}
    \begin{tabular}{|c|c|c|c|}
    \hline
    Component& Quantity & IRCR & FTCP \\
    \hline
    & Accuracy of constituent elements ($\%$) & 98.64 & 99.0 \\
    & MAPE of lattice constants (abc) $(\%)$ & 11.74 & 9.01 \\
    & MAE of lattice constants (abc) (\AA) & 0.67 & - \\
    Decoder & MAPE of lattice angles ($\alpha\beta\gamma$) $(\%)$ & 2.23 & 5.07 \\
    & MAE of lattice angles ($\alpha\beta\gamma$) (degree) & 2.09 & - \\
    & MAE of site fractional coordinates (a.u.) & 0.024 & 0.045 \\
    \hline
    TLB-1 & MAE of $h_\text{form}$ (eV/atom) & 0.162 & 0.051 \\
    & $R^2$ score of $h_\text{form}$ & 0.93 & - \\ 
    \hline
    TLB-2 & MAE of $M_s$ (Tesla) & 0.07 & NA \\
    & $R^2$ score of $M_s$ & 0.87 & NA \\ 
    \hline
    \end{tabular}
    \label{cVAE}
\end{table}

Accuracy of reproducing the constituent elements is $98.64\%$, very close to $99\%$ obtained using FTCP. We use mean absolute percentage error 
(MAPE) and mean absolute error (MAE) as the measures to asses the accuracy of reconstruction of the 
lattice constants and angles. MAE of lattice constants is less than 0.7~\AA~ and that of the lattice angles is only $2^\circ$. 
The site fractional coordinates are reconstructed with a MAE of 0.024. The lattice angles and site fractional coordinates are 
reconstructed much more accurately with IRCR compared to FTCP. 
Only the MAE of lattice constant reconstruction is marginally lower compared to FTCP. 
Overall, the reconstruction performance of the encoder-decoder combination is good, and is comparable to or better than what has been
achieved so far. 

The MAE of $h_\text{form}$ prediction by TLB-1 is 0.162~eV/atom, and the corresponding $R^2$ score is 0.93.
The performance of the TLB for $h_\text{form}$ prediction in Ref.~\cite{FTCP} is somewhat better than our model with a MAE of  0.051~eV/atom. 
One reason for this is perhaps that we have included all materials, stable and unstable, in the training process. In Ref.~\cite{FTCP},
only materials lying up to 80 meV/atom of the convex hull are included.
In our earlier work~\cite{JMMM}, we obtained a MAE of 0.057 eV/atom and $R^2$ score was 0.86. We have to keep in mind that this was a 
relatively simple model trained solely for property prediction, and not for any generative purpose.
For $M_s$ prediction by TLB-2, the MAE is 0.07~T, and the $R^2$ score is 0.87. As a comparison,
we obtained a MAE of 0.15~T and a $R^2$ score of 0.85 for $M_s$ prediction in our earlier work~\cite{JMMM}. So this model
performs quite well.

\begin{figure}
    \centering
    \includegraphics[scale=0.32]{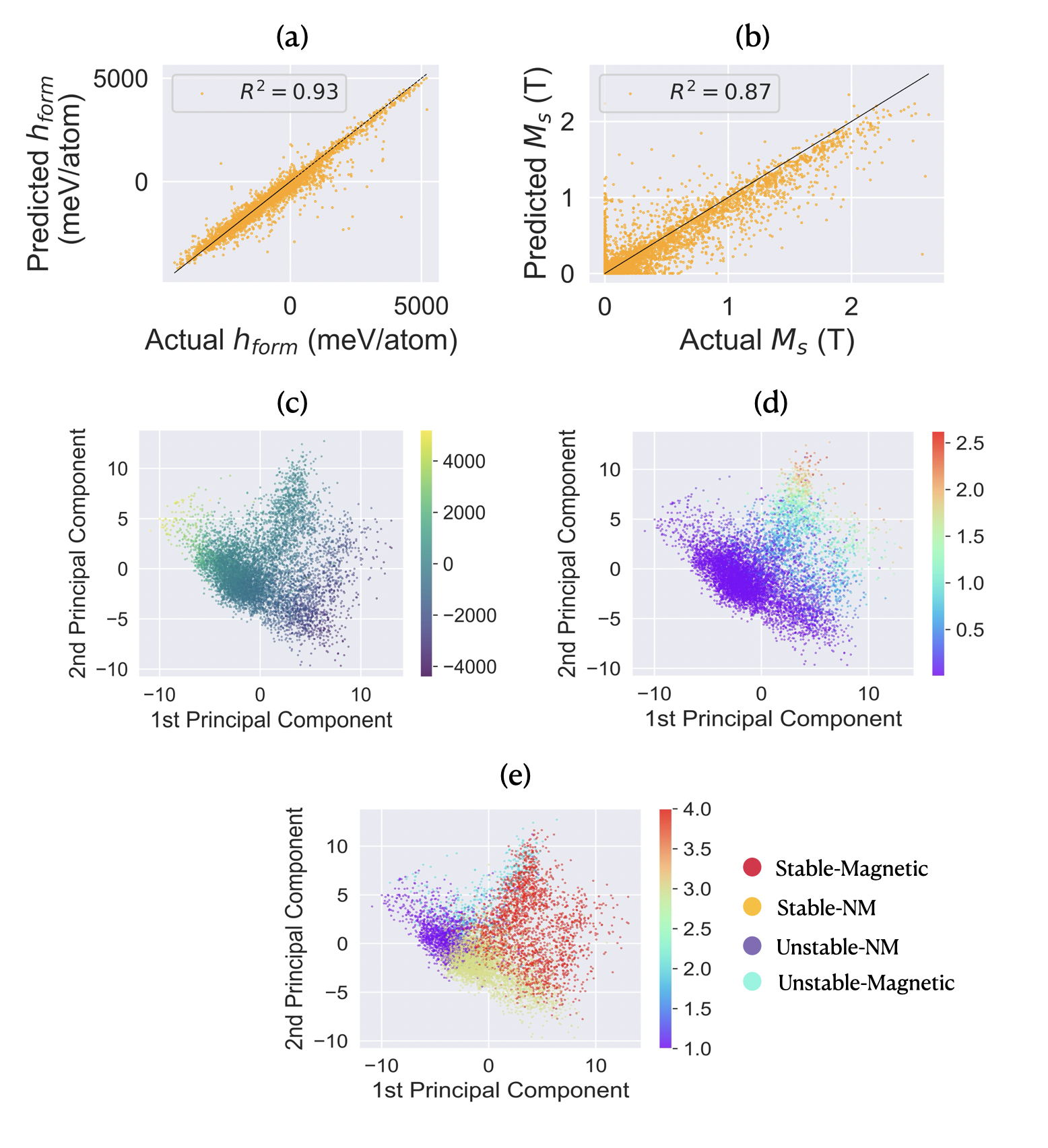}
    \caption{(a)-(b) Parity plots of $h_\text{form}$ and $M_s$ predictions from two target-learning branches, (c)-(e) visualization of two dimensional representation of the latent space, obtained by PCA. Colour schemes are based on $h_\text{form}$ values in meV/atom, $M_s$ values in T, and four material classes, respectively. 
    }
    \label{results}
\end{figure}

We have also shown the parity plots 
(plot of actual vs.\ predicted property values) of $h_\text{form}$ and $M_s$ predictions in Fig.~\ref{results}. 
The parity plot of $h_\text{form}$ demonstrates a balanced prediction over the entire range of values. The plot of $M_s$ shows an overall underestimation 
of $M_s$ by TLB-2, but a slight overestimation at low $M_s$ values. Similar issues with the performance of magnetic moment predictors
have been reported in the literature~\cite{JMMM,JMCC,rhone}. Compared to some of these works, performance of the TLB-2 is better.

Next, we investigate the structure of the latent space to understand how it is organized according to property. 
We take two approaches for this. First, for an easy visualization, we obtain a two-dimensional representation using principal component analysis (PCA). 
PCA is an unsupervised learning model for dimensionality reduction~\cite{tibshirani}. We plot the first two principal components in Fig.~\ref{results}(c)-(e). 
We make two important observations: (1) The points in the latent space are densely packed, centered around the origin as a consequence of minimizing 
the KL loss. (2) There is a continuous change of both $h_\text{form}$ and $M_s$ values indicating property gradient in the latent space. 
In particular, Fig.~\ref{results}(e) suggests that the points in the latent space to which materials get mapped
tend to form four different clusters depending on the two properties. Clusters corresponding to the four possible classes
are color-coded for easy visualization.
These indicate that the latent space of MagGen is a {\em property informed, continuous space}.

Although PCA is successful in providing key insights into the organization of the latent space, 
there must be significant information loss due to dimensionality reduction (from 256 to 2). Hence, in the second approach,
we analyze the structure of the 256 dimensional latent space directly using ideas of graph theory.We believe this is the first attempt at employing graph theoretic tools to efficient materials design.
A {\em simple graph} is a pair $G=(V,E)$ comprising of the sets $V$ and $E$. $V$ is a set of vertices or nodes. 
$E \subset \{\{x,y\} | x,y \in V\; {\rm and} \; x \ne y \}$  is a set of edges. 

In our case, $V$ is the set of all
$\mu$'s to which the training set materials get mapped. Two nodes are connected by an edge if their similarity measure crosses a threshold. 
Tanimoto coefficient, which is found to work well for materials~\cite{tanimoto,cartography}, is used as a measure of similarity. 
Tanimoto coefficient between two nodes $v_1$ and $v_2$, with 
latent space vectors $\vec{z_1}$ and $\vec{z_2}$, is defined as in Eq.~\ref{eq:Tanimoto}.
 
\begin{equation}
   S_{v_1,v_2} = \frac{\vec{z_1}.\vec{z_2}}{ z_1^2 + z_2^2 - \vec{z_1}.\vec{z_2}}.
   \label{eq:Tanimoto}
\end{equation}

\begin{figure}
    \centering
    \includegraphics[scale=0.3]{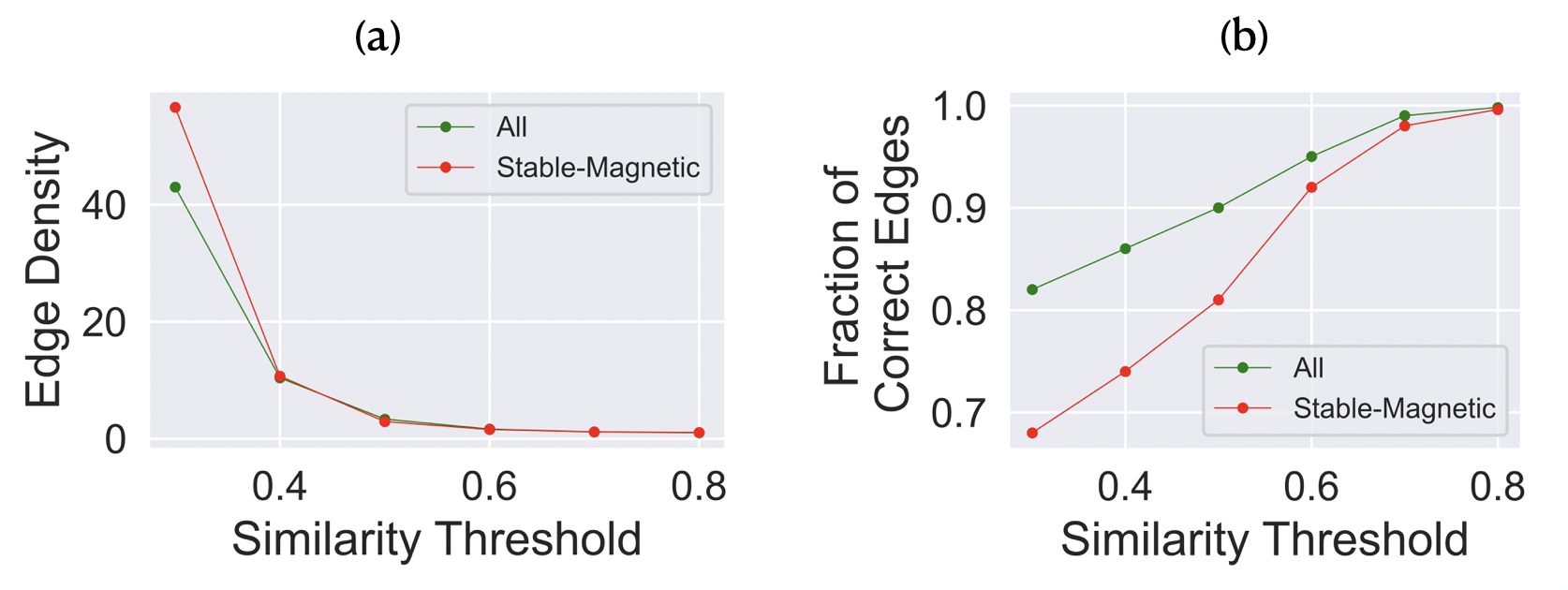}
    \caption{(a) plot of edge density vs similarity threshold, (b) plot of fraction of correct edges vs similarity threshold for all materials and stable-magnetic materials.
    }
    \label{graph}
\end{figure}

Because of property embedding, we expect to find more edges between materials in the same class
compared to materials across classes. 
However, since the model training is a statistical exercise, it is possible to find materials having dissimilar properties
having high similarity. An edge connecting them would be incorrect from materials perspective.   
We consider an edge to be {\em correct} if it connects two nodes representing materials in the same class,
otherwise it is {\em incorrect}. We analyze the structure of the resulting graph for a range of similarity thresholds between 
0.3 and 0.8 using two quantities: (1) {\em edge density}, the number of edges per node, and (2) {\em fraction of correct edges}, the ratio 
of the number of correct edges and the total number of edges. The plots of edge density and fraction of correct edges vs.\ 
similarity threshold are given in Fig.~\ref{graph}(a). Edge density is higher at lower threshold values, as expected.
The graph becomes increasingly sparse with higher threshold values. Fig.~\ref{graph}(b) makes it evident that even for small value of the threshold,
 the fraction of correct edges is quite high. At a threshold of 0.5, this is $90\%$. If we consider only the class of stable-magnetic materials, 
 the fraction of correct edges is greater than $80\%$. This clearly demonstrates that the 256 dimensional latent space of 
 MagGen is organized according to the property values.

With the above insight, in order to generate new stable ferromagnetic materials with $h_\text{form}\le 0$ and $M_s \ge 0.5$~T,
we sample points from the region of the latent space where the materials satisfying the design targets lie. 
By applying the decoder, we obtain the corresponding IRCRs, which are then converted to materials. 
The post-processing steps for converting IRCR to material is discussed in section S3 in SI. 
The two TLBs predict the $h_\text{form}$ and $M_s$ values directly from each sampled point. 

Several methods have been used in the literature to sample points from the latent space, such as local perturbation (Lp)~\cite{FTCP}, spherical linear interpolation (Slerp)~\cite{slerp,FTCP,iMatGen}, and global perturbation (Gp)~\cite{FTCP}.
In terms of validity of the generated crystal structures in the setting of property-driven materials design, Lp works the best~\cite{FTCP}. 
Therefore, we use Lp as the sampling strategy. A schematic diagram of new material generation using Lp is shown in Fig~\ref{schematic}(c).
In this, the latent vector ($\mu$) of a parent material is locally perturbed by adding to it a scaled Gaussian noise, sampled from a standard normal distribution. 
The scale factor controls the trade-off between exploitation and exploration. The generated materials are expected to display high reconstruction accuracy 
with low novelty for a smaller value of the scale factor. For larger values, materials with high novelty are expected, but could have low reconstruction accuracy. 
The effect of scale factor on material generation is discussed in section S4 in SI. Here we have used a scale factor of 1.0. 

We also check the relevance of the graph analysis on the efficiency of materials generation in terms of satisfying the design targets.
First, we generated 15351 materials using Lp around the points representing stable magnetic materials with $M_s\ge0.5$T in the training set without any regards to
how these materials are connected to other materials in the graph. 
Out of these, 11806 materials are predicted to have $h_\text{form}\le0$ and $M_s\ge0.5$~T by the TLBs. 
Thus, $76.9\%$ of the generated materials are predicted to satisfy the design targets. 

Next, we selected parent materials from the training set that not only satisfy the design targets, but also have more than $80\%$ correct edges at a similarity threshold of 0.4. 
We generated a total of 3183 materials using Lp around the points representing the selected parent materials. 
Out of these, 3064 materials are predicted to have $h_\text{form}\le0$ and $M_s\ge0.5$~T by the TLBs. 
Thus, an impressive $96.2\%$ of the generated materials are predicted to satisfy the design targets enabled by property embedding in the latent space. 
While property embedding alone ensures that $\sim 77\%$ of the generated materials satisfy design targets, that number increases greatly
when the connectivity of the parent materials is also taken into account. As a comparison,
Zhao et al.~\cite{CubicGAN} generated 2126042 cubic materials using CubicGAN, and screened them by crystal graph convolutional neural network (CGCNN)~\cite{CGCNN} to predict 
$h_\text{form}$. In absence of property embedding in the latent space only 108897, i.e. only 5\% of the generated materials were predicted to have $h_\text{form}\le0$.

We have also analyzed the structure of the graph after including the 3183 generated materials.
For a similarity threshold of 0.3, $98.1\%$ of the generated materials have more than $90\%$ edges that connect with stable-magnetic materials. 
For a similarity threshold of 0.4,  $74.6\%$ of these materials have all the edges connecting with stable-magnetic materials, and rest of the materials have no edges at all. 
These show that the graph structure of the generated materials strongly correlate with the property predictions from TLBs, and can be used as an additional
check for selecting the promising candidates among the generated materials for further validation.

Finally, we validate the generated materials via DFT. We randomly select 100 materials from the
ones predicted to satisfy both the design targets, and have stoichiometry different from its parent material. We choose this
smaller set since performing complete lattice structure optimization of 3064 materials is beyond our computational resources. 

First, we optimized the lattice shape and size, and the positions of the atoms of the selected materials. 
We then calculated $h_\text{form}$ and $M_s$ in the converged structures. 
Details are given in S1 in SI. 

To assess the performance of MagGen in new materials generation, we use the same three metrics used in Ref.~\cite{FTCP}:
(1) validity rate, defined to be the fraction of materials for which structural relaxation is successful in DFT, 
(2) success rate, the fraction of materials satisfying the design targets as calculated via DFT, 
(3) improvement over random success rate, where random success rate is the probability that a material randomly chosen from the training 
data set satisfies the design target properties, and is approximated by the fraction of materials in the data set that satisfies the design targets. 
Improvement is given by (success rate $-$ random success rate)/random success rate. 

Out of 100 materials, structural relaxation converged for 87 materials giving a validity rate of $87\%$. Among these 87 materials, 
60 materials have $h_\text{form}\le0$ and 68 have $M_s\ge0.5$~T. Overall, 46 materials satisfy both the design targets. 
Thus the success rate for generating materials with $h_\text{form}\le0$ is $60\%$ and that for magnetic materials with $M_s\ge0.5$~T is $68\%$. 
Overall, the success rate of generating materials that satisfy both the design targets is $46\%$. 
In the entire data set, only $16.54\%$ materials have $h_\text{form}\le0$ and $M_s\ge0.5$~T, 
so that the improvement over random success rate is $178\%$.
Ren et al.~\cite{FTCP} obtained validity rates $42.9\%-96.4\%$ and success rates $7.1\%-38.9\%$ for different material design strategies. 
The validity rate of the CubicGAN~\cite{CubicGAN} was only $33.83\%$. 
For PGCGM~\cite{PGCGM}, the validity rate was $93.45\%$ and the success rate was $39.6\%$ for generating materials with  
the sole target of $h_\text{form}\le0$.

 Clearly, MagGen is a significant improvement over the previously reported GMs on a number of points. 
 First, it is trained only on real space 
 features IRCR, which can be directly taken from the materials cif files, and do not require any computations for their construction. 
 Second, by embedding both the target properties
 simultaneously in the latent space, the success rates are improved significantly in a multi-target task, 
 making the materials generation process practically more appealing.  Third, a similarity analysis based on graph
 structure in the latent space helps in generating materials that satisfy design targets more efficiently.  
 Out of these 46, 2 of the most interesting candidate materials (reasons discussed below) are listed in Table~\ref{DFT} along with their DFT calculated properties.
 The complete list of these 46 materials are given in Tables~S5 and S6. 

\begin{table}[ht]
    \centering
     \caption{Composition, $h_\text{form}$, $M_s$, $E_\text{hull}$ and $K_1$ of the 2 candidate stable magnetic materials from MagGen verified by DFT calculations and promising for RE-free permanent magnet applications. 
     }
     \vspace{0.2cm}
    \begin{tabular}{|c|c|c|c|c|}
    \hline
    Composition &  $h_\text{form}$ (eV/atom) & $M_s$(T) & $E_\text{hull}$ (eV/atom) & $K_1$ (MJ/m$^3$) \\
    \hline
    $\text{Fe}_3\text{O}_2\text{F}_2$ & -1.881 & 1.09 & 0.076 & 2.15 \\
    $\text{Mn}\text{Ni}_2\text{O}_2$ & -1.034 & 1.23 & 0.248 & 1.99  \\
    \hline
    \end{tabular}
    \label{DFT}
\end{table}

We wish to add that while calculating the success rate we also considered the materials which have $M_s$ in the range $0.43 \le M_s \le 0.5$~T 
to be successful material generation. The lower bound is obtained by subtracting the MAE of $M_s$ prediction by TLB-2 from our threshold 
0.5~T. This is to explicitly ensure that materials which are within the range of uncertainty of ML prediction are also counted.

In order to have confidence in the ability of MagGen to explore the materials space, we studied how novel the generated materials are. 
We searched for the 46 new magnetic materials in the MP~\cite{MP}, OQMD~\cite{OQMD} and ICSD~\cite{ICSD} databases, and
only 18 of these had entries matching in stoichiometry, but the space groups were different (Table~ S7). 

We calculate $E_\text{hull}$ for the 46 candidate materials for a more complete assessment of thermodynamic stability and experimental synthesizability
using the pymatgen~\cite{pymatgen} PhaseDiagram module, and formation energies of the materials with the same phase in MP.
We found 20 materials having $E_\text{hull}\le0.25$ eV/atom, and six more that are within 0.3~eV/atom of the hull. 
Under appropriate conditions of synthesis, materials that are 300~meV/atom above the hull (1T-MoS$_2$ e.g.) are routinely synthesized.
Therefore, these materials are likely to be experimentally synthesized. 
We want to highlight that only $5.67\%$ of the optimized materials in Zhao et al.'s~\cite{PGCGM} work have $E_\text{hull}\le0.25$~eV/atom 
compared to $20\%$ in our work.

In search of candidate materials for permanent magnets, we calculate $K_1$ for the materials 
which have $M_s\ge1$~T and $E_\text{hull}\le0.25$~eV/atom (Table~S5). Among 12 such materials, 2 materials, ${\rm Fe_3O_2F_2}$ and ${\rm MnNi_2O_2}$,
turn out to have $K_1 > 1$~MJ/m$^3$. Thus, although not explicitly embedded in the latent space, MagGen is able to generate
permanent magnets with large uniaxial anisotropy which are also RE-free. 

There are two minor, but common, shortcomings that MagGen suffers from. 
The discovered materials belong to the space groups ranging from 1 to 8. This suggests that MagGen is biased towards generating low symmetry structures. In fact, other generative frameworks also suffer from a similar 
bias~\cite{iMatGen,CDVAE,superconductor}. Zhao et al~\cite{PGCGM}
added two physics oriented losses based on atomic pairwise distance constraint and structural symmetry that helped generate high symmetry
structures. But no TLBs were used to infuse property information into the latent space. 
Generation of high symmetry structures in property driven materials design is an open question.

IRCR does not have rotational, translational, permutational or supercell invariance argued to be requirement for materials
representations for ML~\cite{sanvito}. 
Most common representations used for generative models do not satisfy full invariance. 
So far, the issue has largely been addressed by data augmentation~\cite{CCCG}. 
Recently Xie et al.~\cite{CDVAE} developed the crystal diffusion variational autoencoder (CDVAE) which ensures full invariance.
CDVAE was used to discover new 2D materials~\cite{Lyngby2022} and next generation superconductors~\cite{superconductor}.
Although it is not clearly understood, experience shows that even without the full invariance properties, representations do lead
to successful materials generation as in voxel image~\cite{iMatGen}, FTCP~\cite{FTCP} and IRCR representation in this work.

On a technical note, we wish to highlight that the success rate of MagGen depends on the reconstruction performance of the decoder and the 
accuracy of property predictions from the two TLBs. Performance of the decoder and the TLBs improve 
rapidly with increasing training data size (Fig.S5). Hence, the success rate can be increased significantly once trained on a 
bigger training data. 

In summary, we used a generative deep learning model to inverse design stable magnetic materials.
DFT calculations were performed to validate the ML predictions. 
We demonstrated that real space features IRCR are sufficient for the inverse design purpose, eliminating  the need of complicated reciprocal space features. 
We created a material latent space that simultaneously encode information about two properties.  
Graph theory was used to analyze the structure of the latent space, and 
this helped us in an ultra-efficient sampling of the space whereby we discovered 46 stable magnetic 
materials by performing DFT on only 100. Overall, we obtained a very high success rate of $46\%$. 
A total of 20 materials are found to be within $0.25$~eV/atom of the respective convex hulls, suggesting the possibility of experimental synthesis.
2 of these are also found to have uniaxial AE in excess of 1~MJ/m$^3$, making them attractive candidates for
rare earth free permanent magnets.

\vspace{0.5cm}
{\bf Acknowledgements:} The work was funded by the DAE, Govt. of India through institutional funding to HRI. All the computations were done on the cluster computing facility at HRI (https://www.hri.res.in/cluster/). SM acknowledges Jyotirmaya Shivottam and Subhankar Mishra for many insightful discussions. 

\newpage

\renewcommand{\thetable}{S\arabic{table}}

\renewcommand{\thefigure}{S\arabic{figure}}

\renewcommand{\thepage}{S\arabic{page}}

\section*{Supplementary Information}

\section*{S1. DFT Calculations}

Spin-polarized DFT calculations are performed within the Vienna ab-initio simulation package (VASP)~\cite{vasp1,vasp2} using a plane-wave basis set. Projector-augmented-wave (PAW)~\cite{PAW} based PBE~\cite{PBE} pseudopotentials are used. $\Gamma$-centered Monkhorst-Pack~\cite{monkhorst} k-point mesh with a density of $0.15~\text{\AA}^{-1}$ is adopted. Energy cut-off for the plane-wave basis set is taken to be the maximum of the default cut-off energies of all the constituent elements in a material as given in the VASP PAW files.

\maketitle
\begin{table}[ht]
    \centering
     \caption{Errors for transition metal atoms in oxides and fluorides}
     \vspace{0.2cm}
    \begin{tabular}{|c|c|}
    \hline
    Species & Error (eV/atom) \\
    \hline
    Mn & 1.668 \\
    Fe & 2.256 \\
    Co & 1.638 \\
    Ni & 2.541 \\
    \hline
    \end{tabular}
    \label{error}
\end{table}

\begin{figure*}
    \centering
    \includegraphics[scale=0.55]{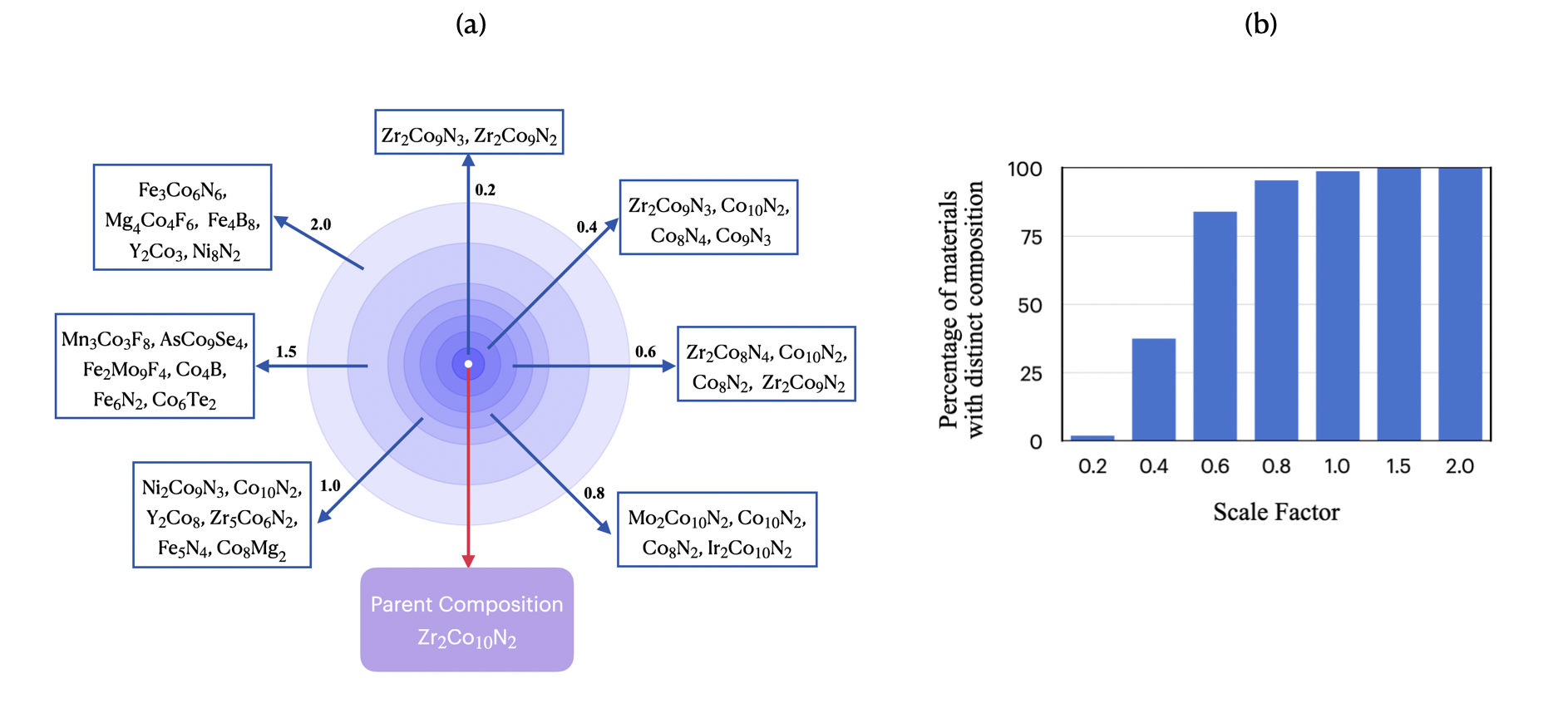}
    \caption{(a) Examples of materials generated by perturbing around Zr$_2$Co$_{10}$N$_2$ with different scale factors; (b) percentage of
    generated materials having different composition than the parent one as a function of scale factor.  
    }
    \label{scale-factor}
\end{figure*}

\begin{figure*}[ht]
    \centering
    \includegraphics[scale=0.45]{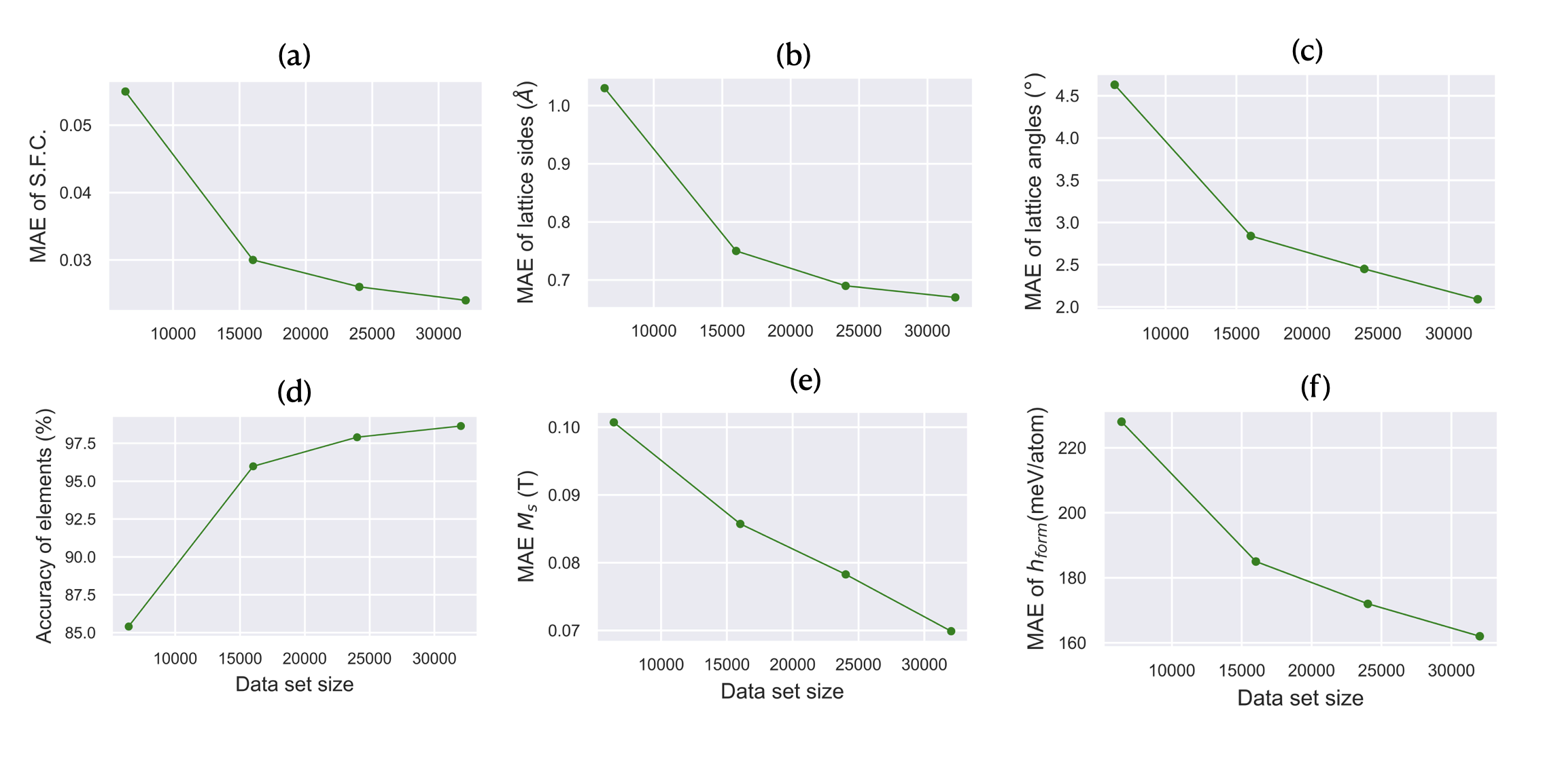}
    \caption{ (a)-(c)Plot of MAE vs training data set size of site fractional coordinates, lattice constants and lattice angles, respectively.(d) Elemental accuracy vs training data set size, (e)-(f) Plots of MAE vs data set size for $M_s$ and $h_\text{form}$ predictions, respectively.   
    }
    \label{train-size}
\end{figure*}

The structures are optimized by changing both the ionic positions and lattice parameters of the simulation cell until the energy and force on each atom converged to less than $10^{-5}$ eV and $10^{-3}$ eV/$\text{\AA}$, respectively. For transition metal oxides and fluorides GGA+U~\cite{dudarev} calculations are done, for the rest of the materials GGA calculations are done. Hubbard `U' values are taken from materials project (MP)~\cite{MP} database (e.g. U for Fe is 5.3 eV). For the calculation of formation energy $h_\text{form}$, GGA/GGA+U mixed approach, proposed by Jain et al.~\cite{PhysRevB.84.045115}, is adopted for transition metal oxides and fluorides, where the total energy of the compound is calculated using GGA+U, but for the reference states GGA calculations are done. The energy corrections for mixing these two frameworks are added according to Eq.\ref{eq:mixing}.  

\begin{equation}
   E_\text{compound}^\text{GGA+U,corrected} = E_\text{compound}^\text{GGA+U} -\sum_M n_M \Delta E_M
   \label{eq:mixing}
\end{equation}
 
Here $M$ represents the transition metals with U value applied and $n_M$ is the number of $M$ atoms present in the compound. The error per $M$ atom is represented by $\Delta E_M$. We have used the same values of $\Delta E_M$ taken from Wang et al.~\cite{Wang2021} work, as used in materials project. The errors are listed in Table~\ref{error}.

Apart from this, anionic energy corrections are also applied~\cite{Wang2021}. Corrections for O species are applied based on local bonding environment: oxide ($d>1.49\text{\AA}$) correction -0.687 eV/atom, peroxide ($1.35 <d<1.49\text{\AA}$) correction -0.465 eV/atom and superoxide ($d<1.35\text{\AA}$) correction -0.161 eV/atom, where $d$ is the nearest-neighbor bond length of O. For F species, the energy correction of -0.462 eV/atom is applied.

For the calculation of anisotropy energy constant $K_1$, a denser k-point mesh with a density of $0.10$~{\AA}$^{-1}$ is used. For $K_1$ calculations in presence of SOC, we increased the energy cut-off by a factor of 1.5. To evaluate $K_1$ of a material, we calculate its total energy for magnetization orienated along the crystallographic $a$, $b$ and $c$ directions. The value of $K_1$ is extracted from these energies using the following relation:
\begin{equation}
E = K_1 \sin^2\theta. 
\end{equation}
where $E$ is the energy `cost' for the spins to rotate away from a particular crystallographic direction by an angle $\theta$.

Materials with $K_1 > 0$ exhibit easy-axis anisotropy, where magnetization prefers to orient along a particular direction.
Materials with $K_1<0$ exhibit easy-plane anisotropy, where magnetization prefers to lie in a plane perpendicular to the crystallographic direction.


\section*{S2. Network Architecture and Hyperparameters}

\subsection*{Encoder}

Encoder consists of four 1D convolution layers with number of filters $\{64,128,256,256\}$, filter sizes $\{5,5,5,3\}$, and strides $\{2,2,2,1\}$. ``SAME'' padding is used at each layer. 
After each convolution layer, LeakyReLU activation function with $\alpha=0.2$ is applied. Batch normalization is applied after that. The output of the last convolution layer is flattened and converted to a latent vector of dimension 256 via two fully connected layers, where the first layer consists of 1024 nodes with sigmoid activation function and the second layer consists of 256 nodes with  linear activation function.

\subsection*{Decoder}

Decoder has the complete mirrored architecture of encoder. It takes the 256 dimensional latent vector as the input and transforms it to a 1024 dimensional vector with a fully connected layer with ReLU activation function. Then the vector is converted to the output shape of last convolution layer of the encoder.
Next four 1D transposed convolution layers are used with number of filters $\{256,256,128,64\}$, filter sizes $\{3,5,5,5\}$, and strides $\{1,2,2,2\}$. ``SAME'' padding is used at each layer. 
Batch normalization and ReLU activation function are applied between the transposed convolutionl layers. The output of the last layer is connected to a 1D convolution layer with number of filters 3 , filter size 3, and stride 1. Finally the reconstructed output is created by passing the output of the last layer through sigmoid activation function.

\subsection*{Target-Learning Branch}

Two target-learning branches are connected to the latent space. Both the branches have same architecture. Each branch takes the mean vector $\mu$ (shape=256) as input. It is then connected to two dense layers with 128 and 32 nodes with ReLU activation function. Then a fully connected output layer with sigmoid activation function produces the predicted target property.

\vspace{0.3cm}

We have optimized the hyperparameters $\alpha_1$, $\alpha_2$, and $\alpha_3$. Optimal values are listed in Table~\ref{hyperparameter}.

\begin{table}
    \centering
     \caption{Optimal values of the hyperparameters}
     \vspace{0.2cm}
    \begin{tabular}{|c|c|}
    \hline
    Hyperparameter & Optimal value \\
    \hline
    $\alpha_1$ & $1e^{-6}$ \\
    $\alpha_2$ & 2 \\
    $\alpha_3$ & 2 \\
    \hline
    \end{tabular}
    \label{hyperparameter}
\end{table}

The model is trained upto 300 epochs using Adam optimizer with a batch size of 256. The learning rate ($lr$) is controlled using a learning rate scheduler to avoid any plateau. The value of $lr$ is given in Eq.\ref{lr}.

\begin{equation}
    lr = 
    \begin{cases}
        5e^{-4}, & \text{epoch} < 100 \\
        1e^{-4}, & 100 \le \text{epoch} < 200 \\
        5e^{-5}, & \text{epoch}\ge 200 
    \end{cases}
    \label{lr}
\end{equation}

The architecture is built using Tensorflow~\cite{Tensorflow} library and all the codes are written in python programming language.


\section*{S3. Post-processing steps for converting IRCR to material}

For converting IRCR to a material cif file, we postprocess its element matrix ($E$) , lattice matrix ($L$), site coordinate matrix ($C$), and site occupancy matrix ($O$), where $E$ and $O$ are one-hot encoded. The lattice constants and angles are obtained immediately from $L$. For $E$ and $O$, the one-hot encoded vectors have been obtained by setting the maximum value in the vector to one and the rest to zero. The unique elements have been then obtained from $E$. We then obtain the valid sites from the matrix $O$. Since there exists one-to-one mapping between $E$ and $O$, the elements which occupy these sites are also obtained. Finally, the fractional coordinates of the valid sites are obtained from respective rows in $C$.


\section*{S4. Effect of scale factor in material generation}

Here we have shown the effect of the scale factor in generating materials from a parent material. We have experimented with seven scale factors having values 0.2, 0.4, 0.6, 0.8, 1.0, 1.5, and 2.0 . We have shown few generated compositions from a single parent composition $\text{Zr}_2\text{Co}_{10}\text{N}_2$ using the above scale factors in Fig.~\ref{scale-factor}(a).
Up to scale factor of 0.6, the generated materials consist of the same elements present in the parent material.
From scale factor of 0.8, elements other than the ones present in parent material begin to appear in the generated materials. 
In Fig.~\ref{scale-factor}(b), we have shown the plot of percentage of generated materials with compositions different from the parent composition 
for different scale factors. With too small a scale factor like 0.2, only $1.95\%$ generated materials have composition different from the parent material. 
For a scale factor of 1.0, $98.82\%$ generated materials have compositions different from the parent material. Beyond this, all generated materials are different in composition
from the parent one. So, more novel materials can be generated using larger scale factors.  

We focus on generating materials with distinct composition, and also look for a higher reconstruction accuracy. To balance the two goals, we have selected the scale factor of 1.0.


\section*{S5. Discovered Materials}

Here we have listed the 46 materials that satisfy both the design targets after DFT calculations.
Among these, 20 materials have $E_\text{hull}\le0.25$ eV/atom, and are listed in Table~\ref{DFT-1}. Rest of the materials have $E_\text{hull}>0.25$ eV/atom, and are listed in Table~\ref{DFT-2}.
Among these materials, 18 have same stoichiometry but different space groups with respect to the materials in MP, OQMD and ICSD databases. These materials are listed in Table~\ref{DFT-3}.

\begin{table*}[ht]
    \centering
     \caption{Composition, space group, $h_\text{form}$, $M_s$, $E_\text{hull}$ and $K_1$ of the 20 candidate stable magnetic materials from MagGen verified by DFT calculations with $E_\text{hull}\le0.25$ eV/atom.
     }
     \vspace{0.2cm}
    \begin{tabular}{|c|c|c|c|c|c|}
    \hline
    Composition & Space Group &  $h_\text{form}$ (eV/atom) & $M_s$(T) & $E_\text{hull}$ (eV/atom) & $K_1$ (MJ/m$^3$) \\
    \hline
    $\text{Fe}_3\text{O}_2\text{F}_2$ & P1 & -1.881 & 1.09 & 0.076 & 2.15 \\
    $\text{Mn}\text{Ni}_2\text{O}_2$ & P1 & -1.034 & 1.23 & 0.248 & 1.99 \\
    $\text{Fe}_6\text{B}_5$ & Cm & -0.167 & 1.294 & 0.192 & 0.80\\
    $\text{Fe}_6\text{B}_4$ & P1 & -0.168 & 1.310 & 0.170 & 0.60\\
    $\text{Fe}_8\text{Ni}_4\text{Si}_4$ & P1 & -0.189 & 1.099 & 0.185 & 0.52\\
    $\text{Mn}_2\text{CoSi}$ & P1 & -0.148 & 1.11 & 0.239 & 0.42\\
    $\text{Fe}_6\text{B}_4$ & P1 & -0.106 & 1.284 & 0.232 & 0.42\\
    $\text{Fe}_8\text{Si}_4$ & P1 & -0.145 & 1.14 & 0.245 & 0.34\\
    $\text{Fe}_8\text{P}_2$ & P1 & -0.107 & 1.704 & 0.234 & 0.17\\
    $\text{Co}_8\text{B}_2$ & P-1 & -0.039 & 1.245 & 0.135 & 0.07\\
    $\text{Mn}_3\text{F}_4$ & P-1 &  -2.421 & 1.39 & 0.123 & -0.02 \\
    $\text{Mn}_4\text{F}_6$ & P1 & -2.528 & 1.27 & 0.143 & -0.01 \\
    $\text{Ti}_3\text{Fe}_4\text{O}_{11}$ & P1 &  -2.425 & 0.78 & 0.150 & - \\
    $\text{Fe}_2\text{O}_2\text{F}_3$ & P1 &  -1.959 & 0.91 & 0.158 & - \\
    $\text{Li}_2\text{Fe}_2\text{F}_4$ & P1 &  -2.376 & 0.92 & 0.179 & - \\
    $\text{Na}_2\text{Fe}_6\text{O}_8$ & P1 &  -1.438 & 0.87 & 0.204 & - \\
    $\text{Zr}\text{Fe}\text{O}_3$ & P1 & -2.640 & 0.48 & 0.241 & - \\
    $\text{Mn}_2\text{CoSi}$ & Pm & -0.186 & 0.96 & 0.201 & -\\
    $\text{Mn}_4\text{Co}_3\text{Si}_6$ & P1 & -0.279 & 0.440 & 0.237 & -\\
    $\text{Mn}\text{Co}_2\text{Si}_2$ & Cm & -0.335 & 0.687 & 0.194 & -\\
    
    \hline
    \end{tabular}
    \label{DFT-1}
\end{table*}

\begin{table*}[ht]
    \centering
     \caption{Composition, space group, $h_\text{form}$, $M_s$ and $E_\text{hull}$ of the 26 candidate stable magnetic materials from MagGen verified by DFT calculations with $E_\text{hull} > 0.25$ eV/atom.
     }
     \vspace{0.2cm}
    \begin{tabular}{|c|c|c|c|c|}
    \hline
    Composition & Space Group &  $h_\text{form}$ (eV/atom) & $M_s$(T) & $E_\text{hull}$ (eV/atom) \\
    \hline
    
    $\text{Mn}_2\text{Se}_3$ & Pm & -0.341 & 0.77 & 0.264 \\
    $\text{Fe}_2\text{F}_3$ & P1 & -2.063 & 1.04 & 0.269\\
    $\text{Mg}_3\text{Mn}_4\text{N}_3$ & Cm & -0.174 & 0.81 & 0.396\\
    $\text{Fe}_6\text{O}_6\text{F}_4$ & P1 & -1.570 & 0.82 & 0.469\\
    $\text{Mg}_2\text{Fe}_4\text{O}_4$ & P1 & -1.274 & 1.45 & 0.540\\
    $\text{Fe}_2\text{Co}_3\text{O}_3$ & P1 & -0.486 & 1.72 & 0.581\\
    $\text{Mn}_2\text{Ni}_2\text{O}_2$ & C2 & -0.725 & 1.66 & 0.594\\
    $\text{Mn}_4\text{S}_5$ & P1 & -0.2 & 0.80 & 0.631\\
    $\text{Ti}_2\text{Mn}\text{O}_2$ & P1 & -1.664 & 0.75 & 0.662\\
    $\text{Co}_5\text{N}\text{F}_4$ & P1 & -0.759 & 1.05 & 0.667\\
    $\text{Fe}_4\text{Co}_2\text{O}_3$ & P1 & -0.287 & 0.92 & 0.716\\
    $\text{Fe}_6\text{N}_6\text{F}_4$ & P1 & -0.360 & 0.59 & 0.769\\
    $\text{Fe}_4\text{F}_2$ & P1 & -0.464 & 1.97 & 0.832\\
    $\text{Fe}_4\text{Co}_4\text{P}_4$ & P1 & -0.254 & 0.76 & 0.375\\
    $\text{Mn}_4\text{Co}_6\text{Si}_4$ & P1 & -0.252 & 0.52 & 0.516\\
    $\text{Mn}_4\text{Co}_5\text{P}_3$ & P1 & -0.258 & 0.66 & 0.296\\
    $\text{Mn}_4\text{Co}_4\text{Si}_4$ & P1 & -0.131 & 0.75 & 0.335\\
    $\text{Mn}_4\text{S}_8$ & P1 & -0.108 & 0.59 & 0.639\\
    $\text{Fe}_6\text{Si}_4$ & P1 & -0.180 & 0.804 & 0.260\\
    $\text{Fe}_6\text{S}_2$ & P1 & -0.052 & 1.550 & 0.278\\
    $\text{Fe}_6\text{P}_4$ & P1 & -0.230 & 0.759 & 0.397\\
    $\text{Fe}_6\text{Co}_5\text{Si}_4$ & P1 & -0.081 & 1.177 & 0.311\\
    $\text{Mn}_4\text{Co}_6\text{Si}_6$ & Cm & -0.216 & 0.640 & 0.289\\
    $\text{Mn}_7\text{Ni}\text{P}_4$ & P1 & -0.270 & 0.563 & 0.337\\
    $\text{Mn}_3\text{Fe}_9\text{S}_6$ & P1 & -0.046 & 0.815 & 0.393\\
    $\text{Zr}_2\text{Co}_6\text{N}_4$ & P1 & -0.086 & 0.461 & 0.544\\
    \hline
    \end{tabular}
    \label{DFT-2}
\end{table*}

\begin{table*}[ht]
    \centering
     \caption{The composition and space group of the 18 materials which had same stoichiometry but different space groups with respect to the materials in MP, OQMD and ICSD databases. 
     }
     \vspace{0.2cm}
\begin{tabular}{ |p{2.5cm}|p{2cm}|p{3cm}|p{3cm}|p{3cm}| }
\hline
Composition & \multicolumn{4}{c|}{Space Group} \\
\cline{2-5} & MagGen & MP & OQMD & ICSD \\
\hline
$\text{Fe}_3\text{O}_2\text{F}_2$ & P1 & P12$_1$/c1, P1m1, Pmma, Pbcm  & - & - \\
\hline
$\text{Fe}_2\text{O}_2\text{F}_3$ & P1 & C12/c1 & - & - \\
\hline
$\text{Na}_2\text{Fe}_6\text{O}_8$ & P1 & R3m & - & - \\ 
\hline
$\text{Zr}\text{Fe}\text{O}_3$ & P1 & - & - & Pbam \\
\hline
$\text{Mn}_2\text{Se}_3$ & Pm & - & C2/m & - \\
\hline
$\text{Fe}_4\text{F}_2$ & P1 & - & Fd$\bar 3$m & - \\
\hline
$\text{Fe}_6\text{S}_2$ & P1 & Pnma, P6$_3$/mmc & p63/mmc, Pm-3m, I4/mmm, Fm-3m & - \\
\hline
$\text{Fe}_8\text{P}_2$ & P1 & Pm$\bar{3}$m & Pmmm & Pmmm \\
\hline
$\text{Fe}_6\text{B}_4$ & P1 & - & R-3c, P4/mbm & - \\
\hline
$\text{Fe}_6\text{P}_4$ & P1 & - & C2/m & - \\
\hline
$\text{Fe}_8\text{Ni}_4\text{Si}_4$ & P1 & F$\bar{4}$3m & F-43m, Fm-3m & - \\
\hline
$\text{Mn}_2\text{Co}\text{Si}$ & P1, Pm & F$\bar{4}$3m & F-43m, Fm-3m & - \\
\hline
$\text{Fe}_4\text{Co}_4\text{P}_4$ & P1 & Pnma & P-62m, Pnma, Amm2, etc. & Pnma \\
\hline
$\text{Mn}_4\text{Co}_4\text{Si}_4$ & P1 & Pnma, F$\bar{4}$3m, P6$_3$/mmc & Pnma, P-62m, P63/mmc etc. & Pnma, P63/mmc \\
\hline
$\text{Co}_8\text{B}_2$ & P-1 & P1m1 & - & - \\
\hline
$\text{Mn}_4\text{S}_8$ & P1 & R$\bar{3}$m,Pnnm, Pa$\bar{3}$ etc. & P21/c, C2/m, P-1, etc.  & Pa-3, Fm-3m \\
\hline
$\text{Fe}_8\text{Si}_4$ & P1 & P$\bar{3}$m1, P4/mmm, I4/mmm  & P-3m1, P-62m, Pnma & P-3m1\\
\hline
\end{tabular}
\label{DFT-3}
\end{table*}


\section*{S6. Effect of training set size}

In this section we have analyzed the performance of MagGen with different training data size given in Fig.~\ref{train-size}.

\clearpage

\bibliography{ref}

\end{document}